\documentclass[conference]{IEEEtran}
\IEEEoverridecommandlockouts

\usepackage{a4wide}
\usepackage{setspace}
\usepackage[dvips]{graphicx}
\usepackage{amsmath, amsthm}
\usepackage{xspace}
\usepackage{amsmath}
\usepackage{amssymb}
\usepackage{amsfonts}
\usepackage{comment}
\usepackage{bm}
\usepackage{comment}
\usepackage{graphics}
\usepackage{listings}
\usepackage{url}
\usepackage{lscape}
\usepackage{mathrsfs}
\usepackage{latexsym}
\usepackage{cite}
\usepackage{verbatim}
\usepackage{url}
\usepackage{mathrsfs}
\usepackage{array}
\usepackage{cases}
\usepackage{multirow}

\DeclareGraphicsExtensions{.eps}

\ifCLASSOPTIONcompsoc
\usepackage[caption=false,font=normalsize,labelfont=sf,textfont=sf]{subfig}
\else
\usepackage[caption=false,font=footnotesize]{subfig}
\fi

\newcommand{\eat}[1]{}
\newtheorem{remark}{Remark}

\begin{document}
\title{Heterogeneous download times in a homogeneous BitTorrent swarm\thanks{This work was supported in part by CAPES, CNPq and FAPERJ (Brazil).}}

\author{\IEEEauthorblockN{
Fabr\'icio Murai, Antonio A de A Rocha\\
Daniel R. Figueiredo and Edmundo de Souza e Silva
}
\IEEEauthorblockA{
\{fabricio, arocha, daniel, edmundo\}@land.ufrj.br \\ \\
COPPE/Systems Engineering and Computer Science Program\\
Federal University of Rio de Janeiro, Brazil
} }

\maketitle

\begin{abstract}

Modeling and understanding BitTorrent (BT) dynamics is a recurrent research topic mainly due to 
its high complexity and tremendous practical efficiency. Over the years, different models 
have uncovered various phenomena exhibited by the system, many of which have direct impact 
on its performance. In this paper we identify and characterize a phenomenon that has not 
been previously observed: homogeneous peers (with respect 
to their upload capacities) experience heterogeneous download rates. The consequences of this 
phenomenon have direct impact on peer and system performance, such as high variability of download 
times, unfairness with respect to peer arrival order, bursty departures and content synchronization. 
Detailed packet-level simulations and prototype-based experiments on the Internet were performed 
to characterize this phenomenon. We also develop a mathematical model that accurately predicts 
the heterogeneous download rates of the homogeneous peers as a function of their content. 
Although this phenomenon is more prevalent in unpopular swarms (very few peers), these 
by far represent the most common type of swarm in BT. 

\end{abstract}

\section{Introduction}

Peer-to-peer (P2P) applications have widely been used for content recovery in Internet. Among 
them, BitTorrent (BT)~\cite{bt} is one of the most popular, used by millions daily 
to retrieve millions of files (movies, TV series, music, etc), accounting for large fractions 
of today's Internet traffic~\cite{urlinternet}. The mainstream success of BT is closely related 
to its performance (e.g., fast download times) and together with its high complexity, has 
triggered the interest of researchers. 

Understanding and characterizing the performance of BT through mathematical models has 
been an active topic of research~\cite{xia_muppala_2010}.
Several studies have uncovered peculiar aspects BT's dynamic, many of which have 
direct impact on system performance. 
Moreover, models that capture user and system performance under 
homogeneous and heterogeneous peer population (with respect to their upload capacities) 
have been proposed for various scenarios \cite{yang_veciana_2004,qiu_srikant_2004,liao_papadopoulos_psounis_2007,chow_golubchik_misra_2009}.
However, most proposed models target large-scale systems, either with a large and 
fixed initial peer population or relatively high arrival rates.

We consider a BT swarm where all peers have identical upload capacities but unconstrained 
(or large) download capacities. In this context, we identify and characterize a phenomenon 
that has not been previously observed: homogeneous peers experience heterogeneous 
download rates. This is surprising because peers are identical and should thus exhibit similar 
average performance and because it has not been captured by any prior model 
(to the best of our knowledge). Moreover, this observation has several important 
implications, such as high variability of download times, unfairness with respect to 
peer arrival order, bursty departures and content synchronization among the peers. 
Two peers are said to be content synchronized after their content become identical at
a given instant. This last consequence is particularly critical since it is closely related to 
the missing piece syndrome \cite{hajek_zhu_2010}.

We characterize the fact that homogeneous peers experience heterogeneous download rates and 
its various consequences by using detailed packet-level simulations and prototype-based 
experiments on the Internet. To underpin critical parameters for this behavior, we consider 
various scenarios. We also develop a mathematical model that explains the phenomenon and predicts 
the heterogeneous download rates of the homogeneous peers as a function of their content. 
The comparison of model predictions with simulation results indicate the model is quite 
accurate. More importantly, the model sheds light on the key insight for this behavior: 
upload capacity allocation of peers in BT depends fundamentally on piece interest relationship, 
which for unpopular swarms can be rather asymmetric.

Finally, the phenomenon we identify is more prevalent in swarms that have a very small 
peer population and usually a single seed (peer with entire content) with limited bandwidth. 
However, this is by far the most prevalent kind of swarm in BT \cite{guo}. Measurement studied 
indicates that more than $35$\% of the swarms have less than $5$ peers at any point in time. 
Thus, we focus our attention on unpopular swarms. 

The rest of this paper is organized as follows. In~\textsection\ref{sec:bt} we present 
a brief overview of BT and motivate the phenomenon we have identified.
In~\textsection\ref{sec:problem} we characterize the phenomenon and its consequences using 
simulations and experiments with a real BT application. \textsection\ref{sec:model} presents 
our mathematical model and its validation. In \textsection\ref{sec:application} we apply 
the model to make predictions about bursty departures. We extend our discussion and 
present some related work in \textsection\ref{sec:disc} and \textsection\ref{sec:related}, 
respectively. Finally, we conclude the paper in~\textsection\ref{sec:conc}.

\section{BT overview and the observed behavior}
\label{sec:bt}

In this section we briefly describe the BT protocol and identify an unexpected behavior common in unpopular swarms.

\subsection{Brief BT overview}

BT is a swarm based file sharing P2P application. Swarm is a set of users
(peers) interested in downloading and/or sharing the same content (a single or a
bundle of files). The content is chopped into pieces (chunks) which are
exchanged among peers connected to the swarm. The entities in a swarm may be of
three different types: (i) the Seeds which are peers that have a complete copy
of the content and are still connected to the system altruistically uploading
data to other peers; (ii) the Leechers which are peers that have not yet fully
recovered the content and are actively downloading and simultaneously uploading
the chunks; and, (iii) the Tracker which is a kind of swarm coordinator, it
keeps track of the leechers and seeds connected to the swarm.

Periodically, the Tracker distributes lists with a random subset of peers
connected to the swarm to promote the interaction among participating peers. In
a first interaction, two peers exchange their bitmaps (a list of all file chunks
they have downloaded). Any latter update in their bitmaps must be
reported by the leecher. 

In order to receive new chunks, the leecher must send ``Interested'' messages to
all peers that announced to have the wanted pieces in their respective bitmaps.
Because of the ``rarest first'' approach specified in BT protocol, leechers
prioritize to download first the chunks that are scarcer in the swarm. Once a
sub-piece of any chunk is received, the ``strict priority'' policy defines that
the remaining sub-pieces from that particular chunk must be requested before
starting the download of any other chunk.

Whenever an ``Interested'' messages is received, peers have to decide whether to
``unchoke'' that leecher and serve the piece or to ``choke'' the peer and ignore
the request. Leechers preferentially upload content to other leechers that
reciprocate likewise, it is based on a ``tit-for-tat'' incentive strategy defined
by BT's protocol. However, a minor fraction of its bandwidth must be dedicated to
altruistically serve leechers that have never reciprocated. This policy,
referred to
as ``optimistic unchoke'', is useful for leechers to boost new reciprocity
relationships. As the seeds do not reciprocate, they adopt the ``optimistic
unchoke'' approach all the time. Those BT policies were designed with the main
purpose of giving all leechers a ``fair share'' of bandwidth. It means that
peers uploading in higher rates will receive in higher download rate, and in a
population of leechers uploading at the same rate, they all must reach equal download rates.

\subsection{The observed behavior}
\label{subsec:strange}

Having presented BT's mechanisms, we now illustrate the heterogeneous download rate 
phenomenon and its consequences with two simple examples. Consider a swarm formed 
by a seed and 5 leechers. All peers, including the single seed, have identical 
upload capacity (64 kBps), but large (unconstrained) download capacity. The leechers 
download a file containing 1000 pieces (256MB) and exit the swarm immediately after 
download completion. The seed never leaves the swarm.
This system was evaluated using a detailed packet-level simulator 
of BT and also an instrumented implementation of BT running on PlanetLab \cite{legout_urvoy-keller_michiardi_2006}.

\begin{figure}[!t]
\begin{center}
\captionsetup[subfloat]{margin=3.2pt}
\subfloat[Arrival intervals: 10s, 10min, 4min, 4min, 4min.]{\label{fig:simul1}
\hspace*{-15pt}
\includegraphics[width=1.6in]{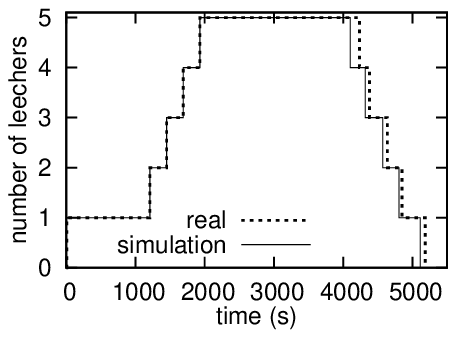}}
\subfloat[Arrival intervals: 10s, 4min, 4min, 4min, 10min.]{\label{fig:simul2}
\hspace*{-15pt}
\includegraphics[width=1.6in]{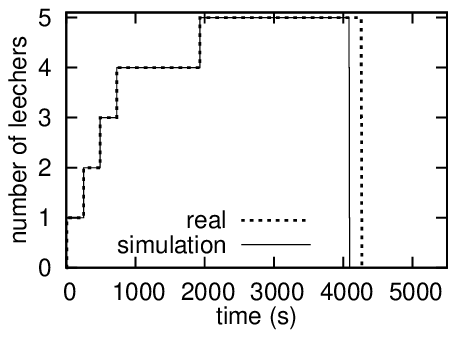}}
\caption{\label{fig:swarm-size} Evolution of the number of leechers in the swarm.}
\end{center}
\end{figure}

Figures \ref{fig:simul1} and \ref{fig:simul2} show the evolution of the swarm size as a function 
of time for both simulation and experimental results and two different leecher
arrival patterns. In Figure \ref{fig:simul1}, peers leave the swarm in the order
they arrived (i.e., FIFO) and have a relatively similar download time. Thus, the
download time is relatively indifferent to arrival order (with the exception of
the first peer).

Figure \ref{fig:simul2} shows the same metric just for different arrival 
times (in fact, the inter-arrival times of peers are also mostly preserved).
Surprisingly, a unexpected behavior can be observed in the system dynamic:
despite the significant difference on arrival times, all five leechers completed
their respective download nearly at the same time.
The time inter departures is small comparing to the download time, which characterizes bursty departures. It means that peers that arrive later to the swarm have a smaller download
time. In fact, the fifth peer completed the download in about half the time of
the first leecher. Thus, the system is quite unfair with respect to the arrival
order of leechers, with late arrivals being significantly favored.
What is happening? Why does BT exhibit such
dynamics? We answer these questions in the next sections.

%
%
%

\section{Heterogeneity in homogeneous BT swarms}
\label{sec:problem}

\begin{figure}[!t]
\begin{center}
\captionsetup[subfloat]{margin=3.2pt}
\subfloat[Corresponding to Figure~\ref{fig:simul1}.]{\label{fig:simul1_downloaded}
\hspace*{-15pt}
\includegraphics[width=1.6in]{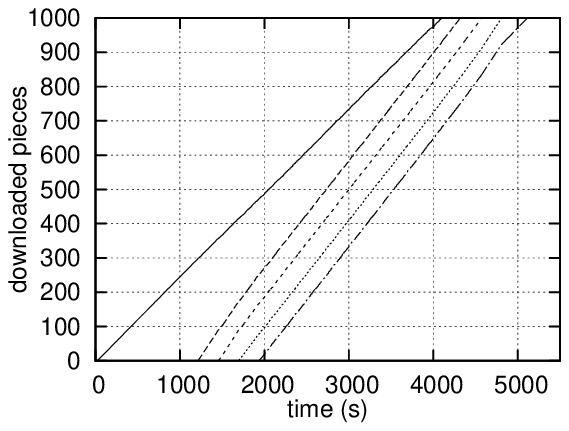}}
\subfloat[Corresponding to Figure~\ref{fig:simul2}.]{\label{fig:simul2_downloaded}
\hspace*{-15pt}
\includegraphics[width=1.6in]{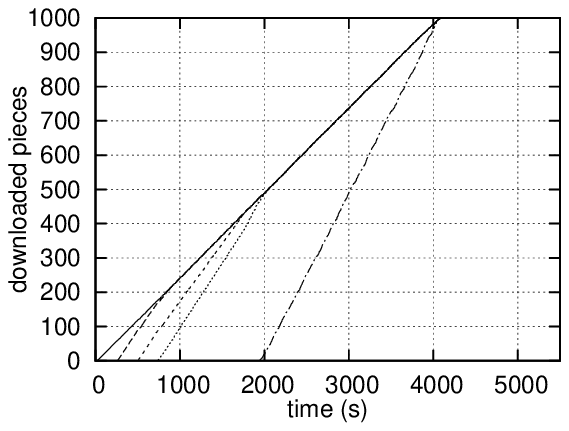}}
\caption{\label{fig:pieces-downloaded} Evolution of the number of downloaded pieces.}
\end{center}
\end{figure}

In order to understand the unexpected behavior exhibited by BT in Figure 
\ref{fig:simul2}, we will analyze the total number 
of pieces each leecher has downloaded over time.
Consider Figures~\ref{fig:simul1_downloaded} and \ref{fig:simul2_downloaded} where each
curve indicates
the total number of pieces downloaded by a given peer for the corresponding
scenario in Figures~\ref{fig:simul1} and \ref{fig:simul2}, respectively. One
can note that the slope of each curve corresponds to respective leecher's
download rate.

We start by considering Figure~\ref{fig:simul1_downloaded}.
Despite the slope of the first leecher being 
smaller than that of the remaining peers, the curves never meet. In particular, 
a leecher finishes the download (and leaves the swarm) before the next 
leecher reaches its number of blocks. We also note that all other leechers 
have very similar slopes. In addition, we observe a peculiar behavior: the slope of 
the fifth leecher suddenly decreases when it becomes the single leecher in the 
system. 

The results illustrated in Figure~\ref{fig:simul2_downloaded} which 
correspond to the scenario considered in Figure~\ref{fig:simul2} show a 
very different behavior. Several
interesting observations can be drawn from this figure. The slope 
of the first peer is practically constant, remaining unchanged by the arrival 
of other peers. The slope of all other peers is larger than that of 
the first peer, meaning the curves may eventually meet. When two curves meet, the 
corresponding  leechers have the same number of blocks and possibly the 
same content (we will comment on this point in the following section). 
The figure also shows that a younger peer does not overcome the first peer, 
but instead the two maintain the same number of downloaded pieces after the
joining point, possibly 
with their contents synchronized. 
Finally, the slope of the second, third and fourth peer are rather similar.
However, the slope 
of the fifth peer is slightly larger than the others, meaning a higher download rate and 
consequently smaller download time. 

%
%
%

In summary, we make the following general observations:
\begin{itemize}
\item 
The first leecher downloads approximately at constant rate.
\item 
Subsequent leechers download at a faster rate than the first.
\item 
Once a leecher reaches the total number of pieces downloaded by the first 
leecher, their download rates are identical.
\item 
Once a leecher reaches the total number of pieces downloaded by the first 
leecher, the download rates of other leechers increase.
\end{itemize}
All these observations are related to the dynamics of BT and will be 
discussed and explained in Section~\ref{sec:model} using a simple mathematical 
model. In the remainder of this section, we discuss the consequences of 
the observed phenomenon and illustrate that it happens even when peer arrival is 
random (i.e., Poisson process).

\subsection{Consequences of heterogeneity in homogeneous swarms}

The observations above imply essentially that the download time of peers are
quite different, despite their homogeneous upload 
capacity. In summary, the consequences are: 
\begin{itemize}
\item
{\bf Variability in download times.} Since peers can experience a 
consistently different download rate, their download times can also differ. 

\item
{\bf Unfairness with respect to peer arrival order.} Since peers download rates, 
and thus download times, may depend on their arrival order, the system is inherently 
unfair, potentially benefiting latecomers in a swarm. 

\item
{\bf Content synchronization.} Due to different download rates and BT's piece 
selection mechanisms (most notably rarest-first), leechers can synchronize on 
the number of pieces they have and, more strongly, on the content itself. This 
means that peers may end up with exactly the same content at some instant,
despite arriving at different points of time.

\item
{\bf Bursty departures.} A direct consequence of content synchronization is bursty 
departures. This means that peers tend to leave the swarm within a small
interval despite arriving at the swarm at relatively far apart instants.

\end{itemize}

Although figures do not show the content synchronization
explicitly, since the first leecher is downloading the file at the same rate at
which the seed push new pieces into the swarm, whenever a leecher reaches the
same number of pieces than it, they have exactly the same content.

Of course, the prevalence of the phenomenon and its consequences depend directly on 
the parameters of the swarm. In particular, the arrival times of peers is certainly 
the most determinant. However, parameters like upload capacity of seed and leechers 
and number of pieces are also fundamentally important. Intuitively, a file with a 
larger number of pieces or a seed with a lower upload capacity increase the
probability that the consequences above occur. In fact, for any arrival 
order of a small set of peers, one can always find system parameters for which
this behavior and its consequences occur.

\subsection{Heterogeneity under Poisson arrivals}

The behavior above does not require deterministic 
arrivals or any crafted leecher arrival pattern. It arises even 
when arrival patterns are random. In this section we characterize the
consequences of the heterogeneous download rates 
phenomenon under Poisson arrivals.

We conducted a large amount of evaluations using detailed packet-level
simulations.
In particular, we consider a BT swarm where 
a single seed is present at all times, while leechers arrive according to 
a Poisson process and depart the swarm as soon as their download is completed. 
In the evaluation that follows, all leechers have the same upload capacity of 
64 kBps (and very large download capacities) and download a file with 
1000 pieces. The upload capacity of the seed ($c_s$) varies between 
48 kBps, 64 kBps, and 96 kBps, and the leecher arrival rate ($\lambda$) is 1/1000. 
These scenarios generate a swarm that has a time average size of 3.7, 3.4 and 3.0 leechers, 
respectively. 

\begin{figure}[!t]
\centering
\includegraphics[width=2.5in]{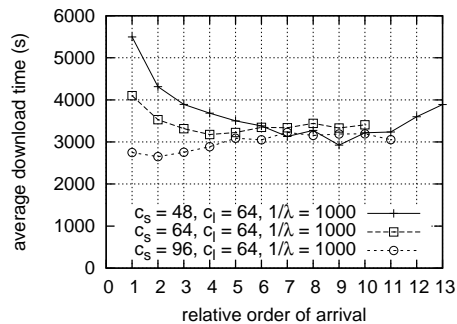}
\caption{Average download time as a function of arrival order in a busy period.}
\label{fig:xx-64_1000_1000_relative-order}
\end{figure}

We start by characterizing the variability in the download times and the 
unfairness with respect to leecher arrival order. Figure \ref{fig:xx-64_1000_1000_relative-order} 
illustrates the average download time for leechers as a function of their
arrival order in a 
busy period. Thus, the $i$-th arrival of a busy period is mapped to index $i$. 
The different curves correspond to different upload capacities of the seed. The 
results clearly indicate that the download time depends on leecher arrival 
order. In particular, for the case $c_s = 64$ kBps, the average download 
time tends to decrease with increasing arrival order, and so the first arrival 
has the largest average download time. Moreover, the download time 
differences are also significant, and can reach up to 30\% (e.g., difference 
between first and fourth arrival).

Figure~\ref{fig:xx-64_1000_1000_relative-order} also indicates that variability in download times strongly 
depends on the seed upload capacity. In particular, a fast seed yields the reverse 
effect: leechers' download times tend to {\it increase} with arrival order. Intuitively, when a 
slow seed is present, late arrivals to a busy period obtain large download rates from other 
leechers, thus exhibiting a lower download time. However, when a fast seed is present, 
the first leecher has the larger upload capacity of the seed until the second arrival, 
thus exhibiting a lower download time. The results also illustrate second order 
effects. For instance, a very late arrival can have an average download time slightly larger 
(or smaller) than a late arrival (e.g., the sixth leecher arrival has longer
download time than fourth for $c_s = 64$ kBps). 
Intuitively, this occurs because a very late arrival is likely to be alone in the busy 
period, having to resort to the seed for finishing the download. Since the upload capacity 
of the seed can be smaller (larger) than the aggregate download rate it receives from 
other leechers, its download time can increase (decrease). This behavior and its
consequences will be explained 
and captured by the mathematical model presented in the next section.

\begin{figure}[!t]
\centering
\includegraphics[width=2.5in]{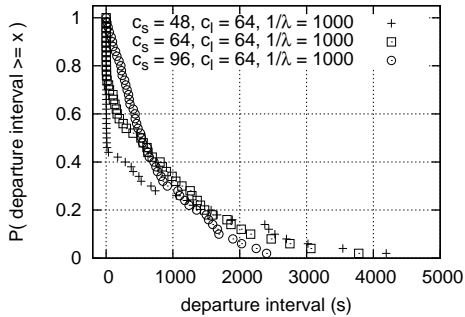}
\caption{Empirical CCDF of the inter-peer departure time conditioned on a busy period.}
\label{fig:xx-64_1000_1000_departure-ccdf}
\end{figure}

In what follows we characterize the burstiness in the leecher departure process.
Figure~\ref{fig:xx-64_1000_1000_departure-ccdf} 
shows the empirical CCDF (Complementary Cumulative Distribution Function) of the leecher 
inter-departure times conditioned on a busy period (i.e., not including the inter-departure 
time between the last leecher in a busy period and the first leecher of the next). Note 
that the peer inter-arrival times follow an exponential distribution with rate 1/1000. However, 
the results indicate a very distinct departure process. In particular, many peers tend to 
leave the swarm at roughly the same time: up to 30\% of peers leave the swarm within a 
couple of seconds from each other (when $c_s = 64$ kBps). Moreover, the departure process 
also exhibits high variability and some peers take as much as ten times more to
leave the system after a departure than the average 
(when $c_s = 64$ kBps). The figure also clearly shows that this observation strongly depends on 
the seed upload capacity, and is more pronounced when the seed is slow. Intuitively, a 
slower seed increases the average download time, thus increasing the chances that 
leechers synchronize their content during the download and depart almost at the same time.
Finally, we also note that a fast seed yields a much less bursty departure process, although 
still favoring shorter inter-departure times.

\begin{figure}[!t]
\centering
\includegraphics[width=2.5in]{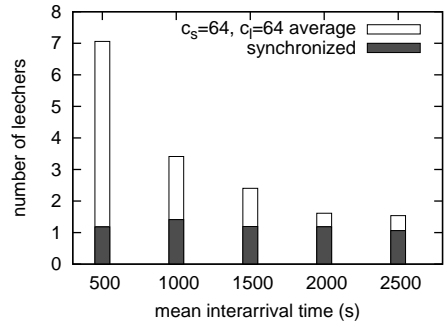}
\caption{Mean number of leechers and mean number of synchronized leechers.}
\label{fig:xx-64_xxxx_1000_synch}
\end{figure}

One consequence of the heterogeneous download rates that is closely related to
the bursty departures is content synchronization.
Figure~\ref{fig:xx-64_xxxx_1000_synch} illustrates the intensity of such
synchronization for different arrival rates. It shows the
average number of leechers in the system and the average number of those which are synchronized.
Here we refer to as synchronized leechers that are not interested in more than 50 pieces
(5\% of the file) of any other. We observe that,
the number of synchronized leechers
remains practically the same as we increase the inter-peer arrival time,
indicating that a larger fraction of peers have similar content when popularity
decreases.

\begin{figure}[!t]
\centering
\includegraphics[width=2.5in]{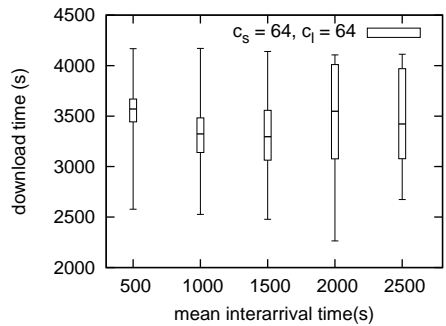}
\caption{Box-plot of download time of leechers for different average inter-arrival times.}
\label{fig:64-64_xxxx_1000_box-plot}
\end{figure}

We next consider the influence of the leecher arrival rate on the download times, 
independently of arrival order. Figure~\ref{fig:64-64_xxxx_1000_box-plot} shows a box plot 
of the download times 
of peers as a function of the average inter-peer arrival time (i.e., the inverse
of arrival rate), for $c_s = 64$ kBps. 
For each scenario, the box plot curve indicates the minimum, 25-th percentile, average, 75-th 
percentile and maximum download times. Note that when the inter-arrival time 
is large (2000 or 2500), the 75-th percentile is very close to the maximum
download time, indicating that many peers have similar download 
times. As the average inter-arrival time decreases, this concentration near the maximum 
diminishes significantly. However, the variability between minimum and maximum
download time does not diminish with the inter-arrival time. In addition, we
run simulations for different values of $c_s$ and observed that a faster seed also 
has strong influence on this behavior, exhibiting a much less concentrated
download times (ommited for conciseness).

\subsection{Real experimental evaluation}

The results shown above were all obtained through simulations but we now 
present results from prototype-based experiments deployed in more realistic 
scenarios. The real experiments were performed in the Internet using machines from Planetlab\cite{planetlab} and running an instrumented version of a BT client\cite{legout_urvoy-keller_michiardi_2006}. Although a large number of experiments were conducted, we 
report only on a limited set of these results due to space constraints. The goal here 
is to validate the phenomenon of heterogeneity in homogeneous BT swarms and its 
consequences in real BT application running over the Internet.  

We consider only private swarms in the experiment, in the sense that only peers 
controlled by the experiment can connect to the swarm for uploading and downloading 
content. Each private swarm consists of a single file of size $S$~MB which is owned by a 
single seed that is always available and has upload capacity of $c_s$. Leechers interested in  downloading the content arrive to the swarm according to a Poisson process with rate $\lambda$. 
All leechers that arrive to the swarm are homogeneous and have upload capacity equal to $c_l$. 
Each experiment run is executed for $t=5,000$~seconds.

\begin{figure}[!t]
\centering
\includegraphics[width=2.3in]{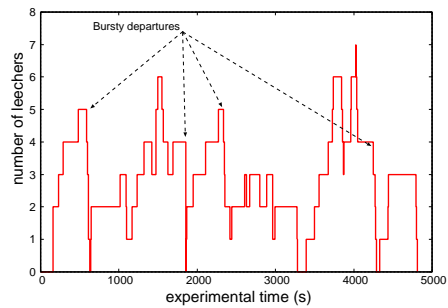}
\caption{Swarm size from real experiment.}
\label{fig:swarm_size_experiment}
\end{figure}

We start by analyzing the evolution of the swarm size for an unpopular swarm. Figure~\ref{fig:swarm_size_experiment} shows the number of leechers in the swarm over time for the 
duration of the experiment, with parameters $\lambda=1/125$ peers/sec., $S=20$~MB, and $c_s=c_l=50$~kBps. We can observe several occurrences of bursty departures, even if leechers arrive according 
to a Poisson process. As previously discussed, bursty departures are consequence of content synchronization among the leechers in the swarm.
 
\begin{figure}[!t]
\centering
\includegraphics[width=2.3in]{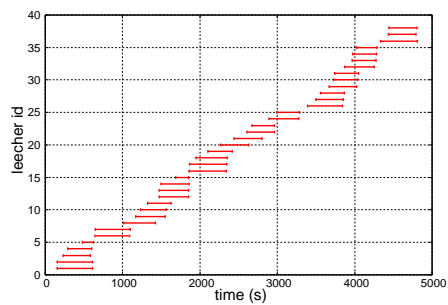}
\caption{Dynamic of the swarm: leechers' arrivals and departures.}
\label{fig:dynamic}
\end{figure}

Using the same experiment as above, we investigate the impact of the leechers' arrival order 
on their download times. Figure~\ref{fig:dynamic} illustrates the dynamics of the swarm, where 
each horizontal line corresponds to the lifetime of a leecher in the swarm, starting when the 
peer arrives and ending when it departs the swarm. Note that peers exhibit significantly 
different download time (which corresponds to their lifetime in the system). In particular, 
in many cases leechers arrive at different time instants but depart in the same burst. 
For instance, the fifth leecher to arrive to the swarm departs in a burst
together with all 
four prior arrivals. Thus, the fifth leecher has a much smaller download completion time, when 
compared to the first leecher. Similar behavior occurs between the fifteenth leecher and the 
three leechers that arrived immediately before. Besides illustrating the variability of the 
download times, this observation also indicates the unfairness with respect to leecher arrival 
order. In particular, late arrivals to a busy period tend to have smaller download times.
 
\begin{figure}[!t]
\begin{center}
\captionsetup[subfloat]{margin=2.3pt}
\subfloat[$c_s=50$~kBps and $c_l=50$~kBps.]{\label{fig:download_ccdf_50}
\includegraphics[width=2.3in]{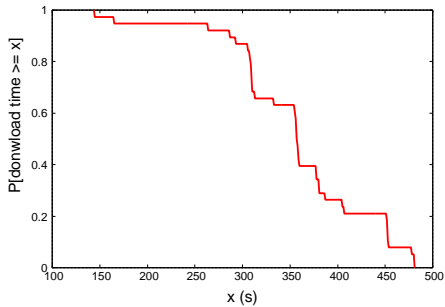}}
\\
\subfloat[$c_s=60$~kBps and $c_l=50$~kBps.]{\label{fig:download_ccdf_60}
\includegraphics[width=2.3in]{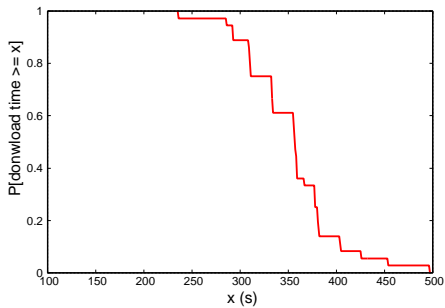}}
\caption{\label{fig:download_ccdf} CCDF of download time from real experiments.}
\end{center}
\end{figure}

We now focus on the distribution of the leechers' download times to illustrate their 
relative high variability. Figures~\ref{fig:download_ccdf_50} and
\ref{fig:download_ccdf_60} show the complementary cumulative distribution function (CCDF) of download times computed for two experiments with 
distinct upload capacities for the seed ($c_s = 50$~kBps and $c_s = 60$~kBps, respectively, 
with all other parameters the same). In both results, download times exhibit a high variance, 
as shown in the figures. In the case $c_s=50$~kBps (Figure~\ref{fig:download_ccdf_50}), the 
minimum and maximum values are 145 and 480 seconds, respectively, with the maximum being more 
than three times the minimum. When the upload capacity of the seed is higher
than that of the 
leechers, Figure~\ref{fig:download_ccdf_60} shows that the variance in download times 
decreases, as expected, since the system capacity is increased. Finally, we note 
several discontinuities (i.e., sharp drops) in both CCDF curves which are caused 
by sets of leechers that have approximately the same download time.

\section{Model}
\label{sec:model}

We develop a simple model attaining to understand the origin of the heterogeneous
download times and its consequences. Our model obtains an approximation to
the average upload and download rates observed by each leecher on different time intervals.


Consider a homogeneous swarm of an unpopular content with a single seed to 
which leechers arrive sequentially and depart as soon as they complete their 
download, such as the one illustrated in Figure~\ref{fig:simul1}. 
In this scenario, bursty departures can only happen if younger leechers
obtain roughly the same number of pieces as older ones, and leave the swarm 
at about the same instant. This in turn implies that younger leechers 
must have higher download rates than older ones, at least for some
periods of time. Why is that? At a given moment, an older leecher $i$ may 
have all pieces owned by a younger leecher $j$. Thus, leecher's $j$ uplink 
capacity will be used by other leechers until $j$ receives a piece 
that $i$ does not have. During this period of time, $j$ simply cannot 
serve $i$, even if it has no other leecher to serve. Therefore, the 
sets of pieces owned by each leecher are the root causes for
heterogeneous download rates and must be considered. 

%

\begin{figure}[!t]
\begin{center}
\captionsetup[subfloat]{margin=3pt}
\subfloat[Leecher $i$ can be represented as server with multiple queues, one for each
neighbor, contaning pieces that are interesting to them.]{\label{fig:multiple-queue}
\includegraphics[width=1.6in]{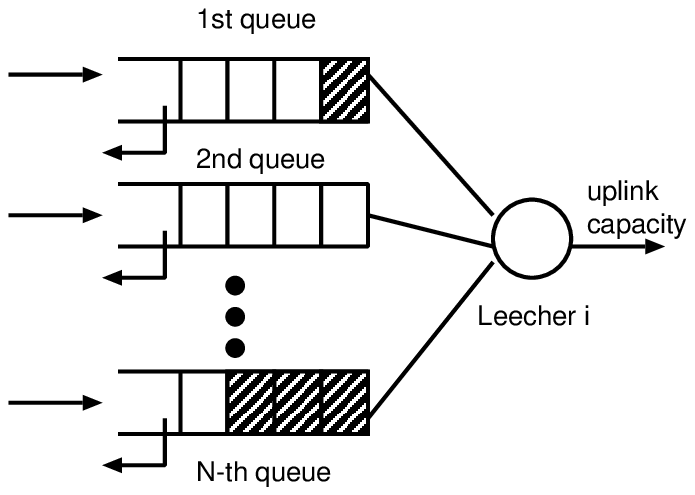}}
\subfloat[The upload bandwidth allocation of leecher $i$ follows a progressive
filling algorithm.]{\label{fig:water-filling}
\includegraphics[width=1.2in]{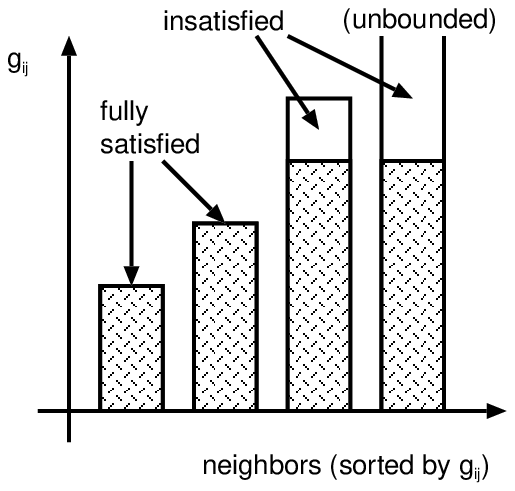}}
\caption{\label{fig:queue-view} Multiple queueing system.}
\end{center}
\end{figure}

In order to capture the observation above, each peer, either a seed or a leecher, 
is represented by a queueing system with multiple queues (see
Figure~\ref{fig:multiple-queue}), one for each 
neighbor, under a processor sharing discipline. 
Queue $i$ of peer $j$ contains the pieces interesting to peer $i$ (i.e., all pieces 
that $j$ has that $i$ has not). When peer $i$ downloads one of these pieces, either 
from $j$ or some other peer, this piece is removed from this queue, as well as all 
other queues where the piece was present. On the other hand, whenever a peer downloads 
a piece that other neighbors are interested in, this piece will be placed in the
queues corresponding to those neighbors, increasing their queues sizes. Finally, 
the queues of the seed always have all pieces that are needed by the leechers. As a 
leecher downloads pieces from the seed and other leechers, this queue decreases, 
eventually becoming empty when the leecher downloads the entire content and departs 
the swarm. We note that the order at which these pieces are served from these queues 
depend on the piece selection policy, but is not important for our discussion. 

Let $c_s$ and $c_l$ be the seed and leechers' uplink capacities, respectively.
Assume that the leechers' downlink capacities are much larger than $c_s$ or
$c_l$. Let $N(t)$ be the number of leechers in the system at time $t$. Since 
the seed always has interesting pieces to every leecher, all the $N(t)$ queues 
in the seed are backlogged. Thus, all queues will be served at rate $c_s/N(t)$. 
Note that, since the swarm is unpopular, we assume the swarm size is small 
enough such that every leecher is neighbor of every other leecher, including the seed. 

A leecher may not have interesting pieces to some of its neighbors at time $t$.
Let a leecher be identified by its arrival order, thus leecher $i$ is the $i$-th
leecher to join the swarm. Also let $n_i(t) \leq N(t)-1$ be the number of
leechers interested in pieces owned by $i$. The instantaneous upload rate 
from $i$ to any of these leechers is $c_l/n_i(t)$.

Whether a leecher has or has not pieces interesting to another
depends on the leechers' respective {\it bitmaps}, i.e. the current 
subsets of pieces owned by a leecher. The set of bitmaps of all leechers 
would precisely determine the exact pieces in each queue. However, 
the dynamics of the bitmaps are intricated and to keep track of them would
be unnecessarily complicated for modeling the phenomenom we are interested in.
Instead, we consider the number of pieces owned by each leecher $i$, $b_i(t),
\forall i$ and infer whether a leecher has interesting pieces to other leechers. 

For the sake of simplicity, let $b_i(t) = b_i$, $N(t) = N$ and $n_i (t) = n, \forall i$. 
Two remarks can be made with respect to $b_i$ and the interest relationship among 
leechers:
\begin{remark}[]
\label{more_pieces}
If $b_i > b_j$, then $i$ has at least $b_i - b_j$ interesting pieces to $j$. 
\end{remark}
\begin{remark}[]
\label{less_pieces}
If $0 < b_i \leq b_j$, it is impossible to determine whether $i$ has or 
has not interesting pieces to $j$ without further information.
\end{remark}

In the following, we will use these two remarks to derive a simple model to 
capture the upload and download rates between the peers. With respect to 
Remark \ref{less_pieces}, we will assume no further information is available, 
and hence the piece interest relationship among peers will be ignored in this 
case.

\subsection{A simple fluid model}

We assume that the content is a fluid, or equivalently, its pieces can be
subdivided in infinitely many parts that can be exchanged (uploaded and downloaded) 
continuously. 


To simplify the explanation, assume that $b_1 > b_2 >
\dots b_N$, i.e. an older leecher has strictly more pieces than a younger 
one. We assume that if leecher $i$ has joined the swarm after $j$, i.e.
$i > j$, $i$ can still upload pieces to $j$ as long as $i$ downloads 
pieces from any peer $k$ that has more pieces than $j$, i.e. $k < j$. 
We also assume that every piece downloaded from the seed by a leecher is 
immediately interesting to all other leechers, independent of their age. 
This assumption is justified due to the rarest first piece selection policy 
used in BT. 


%

Since the seed's upload capacity is $c_s$, each leecher downloads from 
it at rate $c_s/N$. Now let $g_{ij}$ be the rate at which peer $i$
could potentially upload data to peer $j$ provided that there is no capacity 
constraints (i.e. independently of upload and download capacities of 
peers $i$ and $j$, respectively).
If a leecher $i$ is older than $j$, $i$ has interesting pieces to $j$.
Therefore, from the perspective of the multiple queueing system, queue $j$ 
in leecher $i$ is backlogged and $g_{ij} = \infty$. 
On the other hand, if $i$ is younger than $j$, the rate
$g_{ij}$ is given by the rate at which $i$ downloads interesting 
pieces to $j$. According to the previous assumptions, this rate 
is equal to the rate at which peers older than $j$ upload to peer 
$i$. Adding this to the rate at which peer $i$ downloads from the 
seed, we thus have:
\begin{equation}
\label{eq:bandwidth-requirement}
    g_{ij} = c_s/N + \sum_{k<j}u_{ki}, i > j.
\end{equation}
where $u_{ki}$ is the rate at which leecher $k$ uploads to $i$.


We now make an important observation concerning Equation (\ref{eq:bandwidth-requirement}). 
Consider leecher $i$ and some other leecher $j$. The older $j$ is with respect to 
$i$ the smaller is the rate at which $i$ can upload to $j$, that is, the smaller is 
$g_{ij}$. If $j$ is younger than $i$, then $g_{ij} = \infty$. This observation implies 
that $g_{i1} \leq g_{i2} \leq \dots \leq g_{iN}$.

Since the upload capacity of peers is finite, we must now determine 
how the capacity of a given peer $i$ will be divided to serve each of the leechers. 
In particular, recall that $u_{ij}$ is the upload rate from peer $i$ to peer $j$ and note 
that $\sum_k u_{ik} \leq c_l$, where $c_l$ is the upload capacity of a leecher. To determine 
$u_{ij}$ we will use $g_{ij}$ and a bandwidth allocation mechanism that follows a 
progressive filling algorithm, as is illustrated in Figure \ref{fig:water-filling}. 
Roughly, infinitesimal amounts
of bandwidth are allocated to each leecher until no available bandwidth remains
or one or more leechers are satisfied with respect to the $g_{ij} \forall j$ constraints. In 
the latter case, it continues to distribute the capacity among the non-satisfied leechers.
The final bandwidth allocation for leecher $i$ can be obtained by computing the following 
equation in the order $j=1,\dots,N$.
\begin{equation}
\label{eq:model_equations}
u_{ij} = \min\Bigg( g_{ij},
    \frac{ c_l - \sum_{k<j}u_{ik}}
         { n-|\{k|k<j,k\neq i\}| }\Bigg)
\end{equation}
where $|A|$ is the cardinality of a set $A$.
Recall from Equation (\ref{eq:bandwidth-requirement}) that $g_{ij}$
depends on $u_{1,i},u_{2,i},\dots,u_{j-1,i}$, for $i > j$. By calculating
$u_{ij}$ in the order $i=1,\dots,N$, we assure that every variable in
Equation (\ref{eq:model_equations}) has been previously computed.

\begin{figure}[!t]
\centering
\includegraphics[width=1.5in]{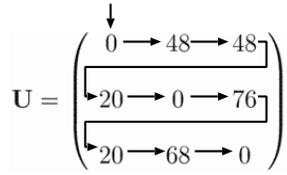}
\caption{Example of matrix $\mathbf{U} = (u_{ij})$ showing the right order of
    calculation.}
\label{fig:matrix}
\end{figure}

As an example, consider the calculation of the matrix $\mathbf{U} = (u_{ij})$, which 
determines upload rates between peers at a given moment, for a small swarm containing 
a single seed and $N = 3$ leechers. Let their upload capacities be equal 
to $c_s=60$ kBps and $c_l=96$ kBps, respectively, and assume $b_1 > b_2 > b_3$.
Matrix $\mathbf{U}$ and the order of computation of its elements are 
depicted in Figure~\ref{fig:matrix}. The download rate for peer $i$ is simply the sum of the elements in column $i$.

Equation (\ref{eq:model_equations}) corroborates the idea that homogeneous 
peers can exhibit heterogeneous upload rates which depend on the number of 
pieces owned by the leechers. Moreover, the younger leechers tend to 
have a higher download rate, as they obtain a higher upload rate from other 
leechers. 

Eventually the number of pieces owned by a leecher may reach the number of 
pieces owned by an older one. In particular, this is bound to occur since 
younger leechers tend to have a higher download rate. In this case, these 
two leechers will no longer have pieces interesting to each other. Thus, 
Equations (\ref{eq:bandwidth-requirement}) and (\ref{eq:model_equations}) 
must be rewritten as functions of $b_i,\forall i$:
\begin{equation}
\label{eq:bandwidth-requirement2}
    g_{ij} = c_s/N + \sum_{b_k>b_j}u_{ki} \; , \, b_i \leq b_j.
\end{equation}
\begin{equation}
\label{eq:model_equations2}
u_{ij} = \min\Bigg( g_{ij},
    \frac{ c_l - \sum_{k|b_k>b_j}u_{ik}}
         { n-|\{k|b_k>b_j,k\neq i\}| }\Bigg)
\end{equation}

Intuitively, Equation~(\ref{eq:model_equations2}) combines the two constraints 
on the rate at which $i$ upload pieces to $j$. The first term stands for the
maximum instantaneous rate irrespective of capacity limitations. The
second term reflects the fraction of $i$'s uplink capacity that can be 
dedicated to $j$ given that some bandwidth has already been allocated. In this 
case, $c_l - \sum_{k|b_k>b_j}u_{ik}$ is the remaining capacity of $i$ 
and $n-|\{k|b_k>b_j,k\neq i\}|$ is the number of peers that will share 
it (including $j$). 

\subsection{Model Validation}
\label{sec:validation}

Our model gives an approximation to the average download rate experienced by 
a leecher in a swarm which depends on the relationship between the number 
of pieces owned by the peers. In this section, we validate the model 
comparing its predictions with simulations results. 

We consider a homogeneous swarm containing $N$ leechers with $c_s = c_l$. 
In this scenario, it is reasonable to assume that 
$b_1 \geq b_2 \geq \dots \geq b_N$ if the index reflects the peer 
arrival order. We partition the set of leechers in two subsets: 
leechers with the same number of pieces as the oldest leecher (subset $A$), 
and those with less pieces than the oldest one (subset $B$). In the scenario 
considered, the model predicts that all leechers in a subset will have 
identical download rates. Moreover, a leecher in $B$ will have a higher 
download rate than one in $A$ and this difference depends on the set 
sizes. In the following, we compare the average download rate of peers 
in each of these sets with simulation results. 

We use deterministic arrivals to reproduce the exact scenarios we intend to
compare. For a swarm with $N$ leechers such that $n_A$ of these belong to 
partition $A$ (i.e. have $b_1$ pieces) the arrivals are set as follows: the 
first $n_A$ arrivals occur next to each other, after they have roughly the 
same number of pieces, i.e., $|b_1-b_i| < 3$, the other $N - n_A$ leechers 
to join the swarm sequentially and far apart. We then compute the average
download rate experienced by a leecher in subset $A$ and for a leecher
in $B$, over a large time interval but before any departures.
\begin{figure*}[!t]
\centerline{\subfloat[Average download rates for leechers in $A$]{\includegraphics[width=2.4in]{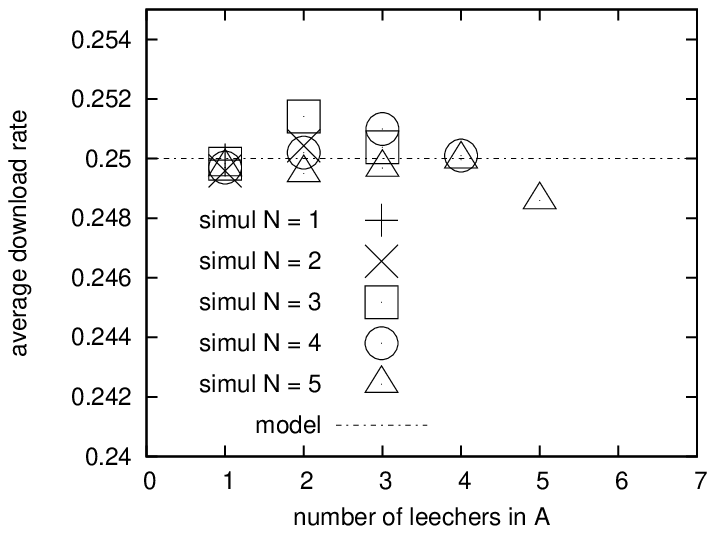}
\label{fig:comparison_A}}
\hfil
\subfloat[Average download rates for leechers in $B$]{\includegraphics[width=2.4in]{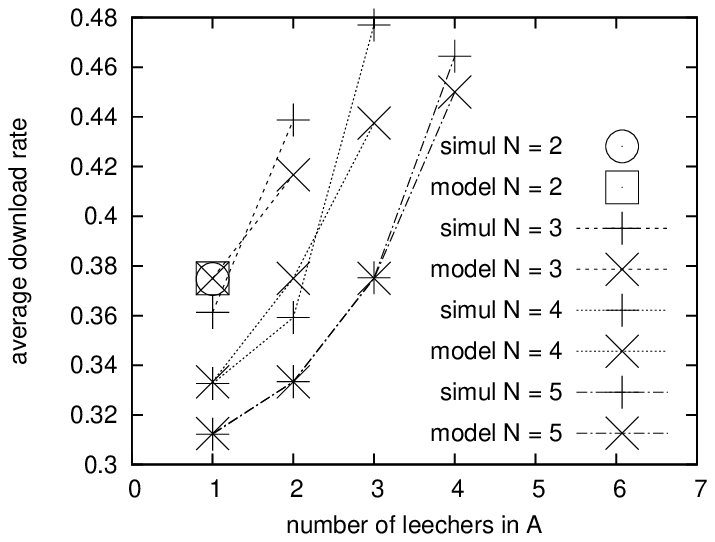}
\label{fig:comparison_B}}}
\vspace*{-2pt}
\caption{Simulation results for $c_s$ = $c_l$ = 0.25.}
\label{fig:comparison}
\vspace*{-12pt}
\end{figure*}

We have simulated 5 runs for each scenario. The confidence intervals obtained are
relatively small and will be omitted. 
The results for $1 \leq N \leq 5$ and $1 \leq n_A \leq N$ are presented in Figures
\ref{fig:comparison}(a,b). Figure \ref{fig:comparison_A} shows simulation and model
results for leechers in $A$. The average download rate of a leecher in $A$
predicted by the model for this scenario does not depend on $N$ or $n_A$ and 
is represented by the horizontal line. Note that model is quite accurate, despite 
the various configurations for $N$ and $n_A$. In particular, the relative error 
is less than 1\% for all scenarios. 

Figure \ref{fig:comparison_B} shows the average download rate for leechers in $B$. 
Since there are numerous points showing either simulation or
model results, we use '+' to identify simulations and 'x' to identify model results
(except for $N=2$, where a circle and a square are used respectively). In
addition, to ease the work of comparing these points, there are lines connecting
results of the same type (simulation or model) for same value of $N$. We note 
that the model is quite accurate, with differences being unnoticeable in many 
scenarios and less than 10\% in all cases. More importantly, the model captures 
well the behavior observed in simulation. For a fixed $N$, as the number of 
leechers in $A$ increases, the average download rate of leechers in $B$ grows. On 
the other hand, for a fixed $n_A$, the average download rate decreases with $N$. Finally, 
a larger number of leechers in the swarm implies a larger range of possible download
rates for leechers in $B$, since $n_A$ can vary from 1 to $N$.

\section{Predicting bursty departures}
\label{sec:application}

The model presented in Section \ref{sec:model} can be used to estimate the 
number of departures that occur in a burst. In particular, consider the 
arrival of a leecher that initiates a busy period (i.e., the first arrival after 
the swarm had no leechers). In the following, we estimate the average number 
of peers that depart the swarm in a burst together with the leecher that 
initiated the busy period.

Let $f$ denote the first leecher of a busy period and assume that the leecher 
arrival follows a Poisson distribution with rate $\lambda$. Also, as assumed by 
the model, a seed is always present and has uplink capacity of $c_s$. Finally, 
let $S$ denote the number of pieces of the content. 

According to the model, the first leecher, $f$, will download the entire content 
at a fixed rate equal to $c_s$, independently on the number of peers in the swarm. 
Note that $c_s$ is also the upper bound on the average download rate, since the seed 
cannot push new pieces into the network at a faster rate. Thus, $f$ will take 
$T = S/c_s$ seconds to finish the download.

Consider arrivals that occur while peer $f$ is in the swarm. The number of such 
arrivals, say $N$, is a random variable and follows the Poisson distribution with 
parameters $\lambda$ and $T$. The download rates of these leechers are a function 
of $N$ and also their instant of arrival. Moreover, as discussed in
Section~\ref{sec:validation}, larger values of $N$ imply a larger spread in the
download rates (see Figure~\ref{fig:comparison_B}). To obtain a conservative lower 
and upper bound on these download rates, we will consider a sufficiently large 
value for $N$. In particular, we use the 99-th percentile of $N$, namely $N_{99}$, and 
thus, $P[N \leq N_{99}] \leq 0.99$.

Given that exactly $N_{99}$ leechers will join the swarm before the departure of 
$f$, we can use the model to obtain the minimum and maximum download rates of these 
peers, independent of their inter-arrival timing. Let $d_{min}$ and $d_{max}$
be, respectively, the minimum and the maximum download rates obtained from the model
given that the swarm has $N_{99}+1$ leechers. Thus, the minimum and maximum time for 
the leechers to obtain the content is, respectively, $S/d_{max}$ and $S/d_{min}$.

Therefore, at least all leechers that arrive before $T - S/d_{min}$ will 
leave the swarm together in a burst with $f$. The expected number of peers that
will arrive within this time period, $B_{min}$ is simply given by 
\begin{equation}
B_{min} = \lambda \, \left( T - \frac{S}{d_{min}} \right)
\label{eq:B_min}
\end{equation}
Similarly, at most all leechers that arrive before $T - S / d_{max}$ will 
leave the swarm in a burst with $f$. The expected number of peers that will arrive 
within this time period, $B_{max}$ is simply given by 
\begin{equation}
B_{max} = \lambda \, \left( T-\frac{S}{d_{max}} \right)
\label{eq:B_max}
\end{equation}
Finally, $B_{min}$ and $B_{max}$ provide a lower and upper bound for the average 
number of leechers that will depart the swarm in a burst with $f$. 

Table~\ref{tab:burst-bounds} shows the expected number of arrivals to the swarm 
before $f$ departs, $E[N]$, which is simply $\lambda T$, and both the lower and 
upper bounds $B_{min}$ and $B_{max}$, respectively. The table shows numerical 
results for different $c_s$ values but with $c_l = 64$ kB/s and $\lambda = 1/1000$.
The results indicate that average number of peers that depart the swarm in a burst 
with $f$ can be significant: between 32\% and 82\% of all arrivals when the seed 
is slower than the leechers and between 10\% and 47\% when they have the same 
upload capacity. We also observe that these ratios reduce as $c_s$ increases, 
indicating that bursty departures are less likely to occur with fast seeds. 

\begin{table}[!t]
\renewcommand{\arraystretch}{1.3}
\caption{Bounds for the expected number of leechers that depart in a burst with $f$, for $\lambda=1/1000$.}
\vspace*{-8pt}
\label{tab:burst-bounds}
\begin{small}
\begin{tabular}{|c|c|c|c|c|c|}
\hline
$c_{s}$ & \multirow{2}{*}{$E[N]$} &  \multirow{2}{*}{$B_{min}$} & \multirow{2}{*}{$B_{max}$}
& \multirow{2}{*}{$\frac{\textrm{$B_{min}$}}{E[N]}$}
& \multirow{2}{*}{$\frac{\textrm{$B_{max}$}}{E[N]}$}\\
\footnotesize{(kB/s)} & & & & &\\
\hline
48 & 5.333 & 1.667 & 4.378 & 0.312 & 0.821\\
\hline
64 & 4.000 & 0.400 & 1.895 & 0.100 & 0.474\\
\hline
96 & 2.667 & 0.000 & 0.857 & 0.000 & 0.322\\
\hline
128 & 2.000 & 0.000 & 0.468 & 0.000 & 0.234\\
\hline
\end{tabular}
\end{small}
\vspace*{-12pt}
\end{table}


\section{General discussions}
\label{sec:disc}

It is interesting to consider the prevalence of the observed phenomenon in more 
general scenarios. Although we have shown its prevalence under a crafted peer 
arrival process and under Poisson arrivals, we claim that homogeneous peers 
can have heterogeneous download rates under very general arrival patterns. 
In particular, given any arrival pattern of peers into a swarm, it 
is possible to choose system parameters (i.e., seed upload capacity, leechers upload capacity, 
and file size) such that the effects described in this paper will be very prevalent. Intuitively,
by choosing a fast enough seed, peers will not be able to disseminate old pieces
before new ones are pushed into the swarm, and thus will have 
significantly different number of blocks, while by choosing a large 
enough file peers are bound to synchronize before they finish the download. In a sense, the 
behavior observed and described in this paper is quite general, although the requirement 
of the swarm being unpopular is important, as we next describe. 

\begin{figure}[!t]
\centering
\includegraphics[width=2.0in]{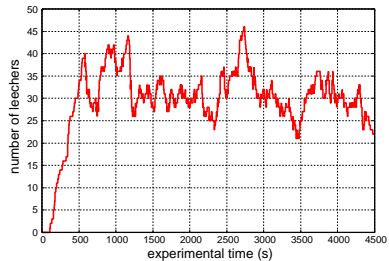}
\vspace*{-10pt}
\caption{Number of leechers of a popular swarm over time ($\lambda = 1/12$, $c_s= 50$ kB/s, $c_l=50$  kB/s)}
\vspace*{-15pt}
\label{fig:popular}
\end{figure}

What happens if we consider very popular swarms, where the peer arrival rate is very large, 
yielding very large swarm sizes? Figure \ref{fig:popular} shows experimental results of the 
dynamics of leecher arrivals and departures for this scenario (Poisson arrivals with 
rate $\lambda = 1/12$ and uplink capacities of $c_s = 50$ kB/s and $c_l = 50$ kB/s). 
Interestingly, we can observe several of the consequences of having heterogeneous 
download rates. In particular, we can observe bursty departures, content synchronization 
and high variability of download times (peers that leave in a large burst have different 
download times, as arrival is well-behaved), for example, at times 600s and 1200s. In a 
sense, the phenomenon is quite prevalent even during the busy period, but not strong 
enough to end the busy period. The characterization and modeling of the phenomenon in 
this scenario is much more entailed, given the complicated dynamics of piece exchange 
of BT and consequently the interest relationship among peers. We leave the investigation 
of these scenarios (popular swarms) as future work. 

Last, we now comment on the relationship of our findings and the {\it missing piece 
syndrome} \cite{hajek_zhu_2010}. The key aspect of this syndrome is content 
synchronization, where a large fraction of peers have all but one and the same 
piece. This situation is particularly bad to the performance of the swarm, as the 
departure rate of the swarm will be equal to the seed upload capacity 
(assuming peers depart as soon as they acquire the last block). Our work has 
shown that peers can synchronize their content much before the last piece. In 
some sense, this generalizes the syndrome to a {\it piece synchronization 
syndrome}, which is inherent to BT dynamics, due to the heterogeneous download 
rates. Once peers have synchronized their content, they can only acquire new pieces 
from the seed, at the upload capacity of the seed. In this situation, 
the {\it missing piece syndrome} is bound to occur.

%
%
%
%
%

\section{Related prior works}
\label{sec:related}


Modeling P2P file sharing systems and in particular BT has been an active area 
of research in the past few years, driven mainly by the high complexity, robustness 
and user-level performance of such systems. One of the first BT models to 
predict the download times of peers was presented in \cite{qiu_srikant_2004}. This 
simple fluid model based on differential equations assumes homogeneous peer 
population (with respect to download and upload capacities) and Poisson arrivals, 
but yields analytical steady state solution. Several subsequent BT models have been 
proposed in the literature to capture various system characteristics, among them 
heterogeneous peer population (with respect to upload and download capacities) 
\cite{piccolo_neglia_2004,liao_papadopoulos_psounis_2007,chow_golubchik_misra_2009}.
BT performance was also studied in the context of
corporate and academic LANs where access links are often symmetric
\cite{meulpolder_2008}.
However, to the best of our knowledge, all models predict that identical peers 
(with respect to their upload capacities) simultaneously downloading a file will
have identical performance (with respect 
to download rates), contrary to the findings in this paper. Moreover, BT models 
generally assume either a rather large peer arrival rate (e.g., Poisson) or a large 
flash crowd (all peers join the swarm at the same time). This is somewhat surprising, 
given that most real BT swarms are rather small in size and quite unpopular
\cite{guo}. Finally, one perverse effect of this lack of popularity, content
unavailability, is shown to be a severe problem found in most of BT swarms~\cite{conext2009}.

Another interesting aspect of BT has been the discovery and characterization of 
some non-trivial phenomena induced by its complex dynamics. For example, peers in BT 
swarm tend to form clusters based on their upload link capacities, exhibiting a strong 
homophily effect. In particular, peers with identical upload capacities tend to exchange 
relatively more data between them \cite{legout_liogkas_kohler_zhang_2007,bharambe_herley_padmanabhan_2006}. 
Another interesting observed behavior is the fact that arriving leechers can continue to 
download the entire content despite the presence of any seed in the swarm, a property 
known as self-sustainability \cite{menasche_et_al_2010}.
More recently, a phenomenon known as {\it missing piece syndrome} has been identified and 
characterized mathematically, which states that in large swarms of long durations, the 
system can become unstable (i.e., number of leechers diverges to infinity) if the upload 
capacity of the seed is not large enough \cite{hajek_zhu_2010}. This last phenomenon is 
quite related to our work and was discussed in Section \ref{sec:disc}. Again, to the best 
of our knowledge, we are not aware of any prior work that has alluded the phenomenon 
we describe in this paper, namely, that homogeneous peers can have heterogeneous 
download rates.

\section{Conclusion}
\label{sec:conc}

This paper identifies, characterizes and models an interesting phenomenon in 
BT: Homogeneous peers (with respect to their upload capacity) experience 
heterogeneous download rates. The phenomenon is more pronounced in 
unpopular swarms (few leechers) and has important consequences that directly 
impact peer and system performance. The mathematical model proposed captures 
well these heterogeneous download rates of peers and 
provides fundamental insights into the root cause of the phenomenon. Namely, 
the allocation of system capacity (aggregate uplink capacity of all peers) among 
leechers depend on the piece interest relationship among peers, which for 
unpopular swarms is directly related to arrival order and can be significantly 
different.

\bibliographystyle{IEEEtran}
\bibliography{SyncBT}

\end{document}